\documentclass[11pt]{article}
\newlength{\vshift}
\newlength{\hshift}
\usepackage{amssymb}
\usepackage{amsmath}
\usepackage{amsthm}
\usepackage{amsopn}

\def\nn{\nonumber }
\def\la{\lambda}
\def\La{\Lambda}

\def\ka{\kappa}
\def\de{\delta}
\def\be{\beta}
\def\ga{\gamma}
\def\al{\alpha}
\def\si{\sigma}

\def\shouldid{\stackrel{!}{=}}

\def\ds{\stackrel{\star}{,}}
\def\x{\hat x}

\def\p{\partial}

\def\lb{\lbrack}
\def\rb{\rbrack}
\def\pat{\partial}

\providecommand{\href}[2]{#2}

\begin{document}

\begin{titlepage}
\rightline{LMU-TPW 2004-07}
\rightline{MPP-2004-110}

\vspace{4em}
\begin{center}

{\Large{\bf Second order expansion of action functionals\\}}\vspace{1em}{\Large{\bf of noncommutative gauge theories}}

\vskip 3em

{{\bf  Lutz M\"oller}}

\vskip 1em

Universit\"at M\"unchen, Fakult\"at f\"ur Physik\\
        Theresienstr.\ 37, D-80333 M\"unchen\\[1em]

Max-Planck-Institut f\"ur Physik\\
        F\"ohringer Ring 6, D-80805 M\"unchen\\[1em]
 \end{center}

\vspace{2em}

\begin{abstract}
Field theory and gauge theory  on noncommutative spaces have been
established as their  own areas of
research in recent years. The hope prevails that a
noncommutative gauge theory will deliver testable experimental
predictions and will thus be a serious candidate for an extension of
the Standard Model. This note contains the results for expanded gauge
theory actions on a noncommutative space with constant $\theta^{\mu\nu}$, up to
second order, together with a discussion of the ambiguities
of the expanded theory 
and how they affect the action.
\end{abstract}
\vskip 0.5cm
\qquad\hspace{2mm}\scriptsize{eMail: lmoeller@theorie.physik.uni-muenchen.de}
\vfill

\end{titlepage}\vskip.2cm

\newpage
\setcounter{page}{1}

\section{Introduction}
\label{A}
Noncommutative (NC) gauge theory has been an intensively
discussed issue during the past five years, prompted by  the results of Seiberg
and Witten's seminal paper \cite{sw}, connecting commutative and NC
gauge theory (on the canonical NC space, see below). Most of the work on this subject covered the
all orders or summed-up NC gauge theory, typically discussed in the
string theory context. Recent results \cite{chai04}, \cite{wess04} show that NC spacetime might 
be endowed with a deformed symmetry structure. Still, the phenomenon of so-called
ultraviolet-infrared mixing \cite{uvir1}, \cite{uvir2}, \cite{armoni1}  causes concerns to treat the
summed-up NC field theory as a serious candidate for a realistic
extension of the Standard Model of elementary particle physics. 

It has been shown in a series of papers \cite{mssw}, \cite{jssw},
\cite{jmssw} that the results of \cite{sw} can also be obtained in a setting entirely independent
of string theory. The ansatz here uses only algebraic properties of
the canonical NC space via the properties of the
$\star$-product. The canonical NC space is characterised by noncommuting
coordinate functions $\lb \x^\mu,\x^\nu\rb=\theta^{\mu\nu}$ with a
constant background field $\theta^{\mu\nu}$. This NC space can
be realised in terms of ordinary functions by means of the
$\star$-product $(f\star
g)(x)=\exp(\frac{i\hbar}{2}\theta^{\mu\nu}\frac{\p}{\p
  y^\mu}\frac{\p}{\p z^\nu})f(y)g(z)|_{y,z\rightarrow x}$.
 
The most important result of \cite{jmssw} is that such a purely constructive ansatz works. The
$\star$-product is treated in this ansatz in an expanded way
(partially up to second order in \cite{jmssw}). Because of the
expanded approach the
conceptual problems related to the UV/IR mixing are avoided, since an
effective IR-regulator is introduced. The
expanded approach introduces and uses enveloping algebra-valued gauge
theory. Therefore no
restrictions exist on the admissible  gauge groups, especially $SU(N)$-gauge
theory can be realised on the NC space. Thus, a NC
generalisation of the Standard Model has been constructed
\cite{cjssw}. While no statements
about uniqueness or existence can be made a priori, the
approach can be formalised and 
existence can be proven \cite{quadri}. 

Several authors have analysed such an expanded version of a NC Standard Model
with a view to derive phenomenological consequences. The important observation  of \cite{cjssw}
has been that the expanded NC Standard Model does not suffer from the physically problematic charge quantisation 
that had been observed in the non-expanded NC gauge theory \cite{hayakawa1}, \cite{hayakawa2}. 
Also the tensor product of several gauge groups, necessary for a generalisation of the Standard Model, 
can be treated in a satisfactory way in this context. Again, this issue had been highly non-trivial in the 
non-expanded approach \cite{chaichian2}.

The key phenomenological result is that the expanded approach leads to new nonlinear couplings between
the gauge bosons of the different gauge sectors of the Standard Model. 
Also matter fields acquire new multiple couplings to gauge bosons that are classically prohibited, 
e.g. neutral particles may couple to the photon because of the NC geometry. Therefore this model predicts higher-dimensional operators 
which are clearly power-counting non-renormalisable. This is acceptable, if the theory is regarded as an
effective physical theory, presupposing other physics at some higher scale. 

Some phenomenological studies, e.g.
\cite{rizzo}, treat $\theta^{\mu\nu}$ as an ether-like Lorentz invariance violating field pervading
spacetime. The calculated experimental bounds are very high. In contrast, $\theta^{\mu\nu}$-expanded models should be 
regarded as IR-regularisable. Two sorts of phenomenological consequences can be distinguished: 
In \cite{expe1} and \cite{expe4} the pure gauge sector has been investigated,
 Standard
Model-forbidden vertices of three (possibly different) gauge bosons have been found. This means that the Z boson could decay into two $\ga$ or two
gluons, which might well be measurable. However, due to an ambiguity in the choice of the matrix trace the pure gauge sector of NC gauge theory is not very
well suited to derive strict bounds on the NC scale. Ongoing
analysis manages to constrain the trace in
the electroweak sector \cite{expe4}, still this freedom limits predictiveness. Therefore, much energy has been devoted as well to studies
of $\theta^{\mu\nu}$-expanded NC effects in the fermionic sector
\cite{carlson},  \cite{carroll}, \cite{expe2} and \cite{expe3}. Yet, as has been argued already  \cite{calmet2004}, great care has to be taken to evaluate the bounds on 
the NC scale by such analyses. 

In any case, all phenomenological analyses center on effects deriving from first-order terms in  $\theta^{\mu\nu}$. The concise calculation of second order
terms in  $\theta^{\mu\nu}$ has so far been ignored due to the technical complexity of some calculations.
This work fills this gap. We recall the outline of the basic approach first developed in
\cite{jmssw} and proceed to presenting full results up to second order in
$\theta^{\mu\nu}$ in the action functionals. 

In the meantime, many
discussions have continued the work started in \cite{jmssw}, e.g. the
quantisation of this expanded
model has been discussed \cite{bichl}, \cite{grimstrup}. 
We choose among the possible actions for the NC gauge theory those which are direct generalisations of the actions
of the Standard Model of particle physics. We will omit the discussion
of other interesting 
actions such as Born-Infeld  (but the approach
would be analogous).  

With this restriction to Standard Model-type
actions we could of course miss out on models which might be more
suited to NC spaces, cp. \cite{grosse2003}. Also we omit pure scalar field theory, the
Einstein-Hilbert action and supersymmetric actions. These models
have been discussed by now  in the literature, e.g. \cite{chamseddine},
\cite{mikulovic}. We have also applied this approach to other NC
spaces, such as the $\ka$-Minkowski spacetime \cite{f2}.

%
%
%

\section{Enveloping algebra-valued gauge theories}
\label{construction}

We start by fixing the notations for infinitesimal gauge transformations on a
commutative space \cite{yangmills}: 
\begin{equation}
\label{ga30}
\de_\al \psi^0_i(x) =i \al_a(x) (T^a)_{ij} \psi^0_j(x).
\end{equation}
The field $\psi^0(x)$ is in the fundamental representation of an
arbitrary non-Abelian  gauge
group. Abelian simplifications are not spelt out.  
We can considerably simplify the notation of (\ref{ga30}), keeping the generators of the gauge
Lie algebra and the $x$-dependence of  $\psi^0$ and  $\al$ implicit, $\al\equiv \al_a(x)T^a$ and
$\psi^0\equiv\psi^0(x)$:
\begin{equation}
\label{ga32}
\de_\al \psi^0 =i \al\psi^0.
\end{equation}
Since the generators $T^a$ form a Lie algebra $\lb T^a, T^b\rb =if^{ab}_c T^c$, the commutator of two infinitesimal
gauge transformations closes:
\begin{equation}
\label{ga32a}
(\de_\al\de_\be -\de_\be\de_\al) \psi^0 =\al\be
\psi^0-\be\al\psi^0=\de_{-i\lb\al,\be\rb} \psi^0.
\end{equation}

A gauge transformation acts on the
conjugate transpose of a field 
$\de_\al \overline{\psi^0} =-i \overline{\psi^0}\al$, such that
$\de_\al (\overline{\psi^0}\psi^0) = 0$.
Derivatives $\pat_\mu\psi^0$ can be rendered gauge
covariant by adding a gauge
potential:
\begin{equation}
\label{ga34}
\de_\al (\mathcal{D}^0_\mu \psi^0)= \de_\al \big((\pat_\mu-iA^0_\mu)
\psi^0\big) \shouldid i \al(\mathcal{D}^0_\mu \psi^0),\hspace{4mm} 
\Rightarrow \hspace{4mm} \de_\al A^0_\mu =\pat_\mu \al -i \lb A^0_\mu, \al\rb.
\end{equation}
The gauge potential transforms in the adjoint representation, as does the
field strength $F^0_{\mu\nu}$, which is constructed from the
commutator of two covariant derivatives
\begin{eqnarray}
\label{ga35}
F^0_{\mu\nu}=i\lb\mathcal{D}^0_\mu, \mathcal{D}^0_\nu\rb=\pat_\mu
A^0_\nu-\pat_\nu A^0_\mu-i\lb A^0_\mu, A^0_\nu\rb,\hspace{4mm}
\Rightarrow \hspace{4mm} \de_\al F^0_{\mu\nu}& =&
i\lb\al,F^0_{\mu\nu}\rb,\\
\lb\mathcal{D}^0_\la,F^0_{\mu\nu}\rb+\lb\mathcal{D}^0_\mu,F^0_{\nu\la}\rb+\lb\mathcal{D}^0_\nu,F^0_{\la\mu}\rb&=&0 \quad \textrm{(Bianchi identity)}. \nn
\end{eqnarray}

In analogy, we start the analysis of gauge theory on  NC space
based on  infinitesimal gauge transformations. The NC space is represented on the algebra of functions of
commutative variables by a $\star$-product. Therefore  (\ref{ga30}) is
replaced by\footnote{For convenience we choose the less correct notation $\La(x)\star
  \psi(x)$ instead of $(\La\star\psi)(x)$.}
\begin{equation}
\label{ga36}
\de_\La \psi(x) =i \La(x) \star \psi(x).
\end{equation} 

As before, the gauge transformation of the field $\psi(x)$ is implemented by left
$\star$-multiplication of $\psi(x)$ with a function $\La(x)$. This is
again a local gauge transformation, although this statement has to be taken with a grain of
salt, since the $\star$-product involves an arbitrary number of
derivatives and is therefore highly non-local. We will use the
$\star$-product always in an expanded form, thus it is local at every
order of the expansion (this does not imply
all-orders locality).
Representations for which a field $\psi(x)$ is multiplied
from the right are possible as well.
The other quantities of NC gauge theory are defined in the following way:
Since the Moyal-Weyl $\star$-product is $x$-independent, it commutes
with derivatives. Therefore we define  on the canonically NC space  as
generalisations of (\ref{ga34}) and (\ref{ga35}) a NC covariant
derivative along with a NC
gauge potential and the NC field strength:
\begin{eqnarray}
\label{ga40}
\de_\La (\mathcal{D}_\mu \psi)= \de_\La \big(\pat_\mu\psi-iA_\mu\star
\psi\big) \shouldid i \La\star\mathcal{D}_\mu \psi,\hspace{4mm}
\Rightarrow \hspace{4mm} \de_\La A_\mu &=&\pat_\mu \La -i \lb A_\mu\ds \La\rb\nn\\
F_{\mu\nu}=i\lb\mathcal{D}_\mu\ds \mathcal{D}_\nu\rb=\pat_\mu
A_\nu-\pat_\nu A_\mu-i\lb A_\mu\ds A_\nu\rb,\hspace{4mm}
\Rightarrow \hspace{4mm} \de_\La F_{\mu\nu}& =
&i\lb\La\ds F_{\mu\nu}\rb,\\
\lb\mathcal{D}_\la\ds F_{\mu\nu}\rb+\lb\mathcal{D}_\mu\ds F_{\nu\la}\rb+\lb\mathcal{D}_\nu\ds F_{\la\mu}\rb&=&0 \hspace{2mm} \textrm{(Bianchi identity)}. \nn
\end{eqnarray}

For a
 Lie algebra-valued gauge parameter $\La(x)=\La_{a}(x)T^a$, two gauge transformations in general
do not close anymore as in (\ref{ga32a})
\begin{eqnarray}
\label{ga38}
(\de_{\La_1} \de_{\La_2} -\de_{\La_2} \de_{\La_1})\psi(x) & =&
\frac{1}{2}\lb T^a,T^b\rb\{\La_{1,a}(x)\ds\La_{2,b}(x)\} \star
\psi(x)\\
&&+\frac{1}{2}\{ T^a,T^b\}\lb\La_{1,a}(x)\ds\La_{2,b}(x)\rb \star
\psi(x)
.\nn
\end{eqnarray}
Only a $U(N)$ gauge theory allows to express the anti-commutator  $\{
T^a,T^b\}$  again in terms of the generators $T^a$ \cite{sw}.
The only alternative
is that the concept of Lie algebra gauge theories has to be
generalised, towards the enveloping algebra of the Lie algebra we
started with. The
 Lie algebra and its enveloping algebra are the only mathematical entities,
 which are consistent with $\lb T^a, T^b\rb =if^{ab}_c T^c$, independent of a specific
 representation. The enveloping algebra $\mathcal{A}_T$ of the Lie
 algebra is an infinite-dimensional algebra freely generated by $T$
 and divided by the ideal generated by $\lb T^a, T^b\rb =if^{ab}_c T^c$
 \cite{jssw}. It consists of all (symmetrically) ordered tensor powers
 of the generators $T^a$ and therefore is an infinite-dimensional tensor
 algebra. 

Since the anti-commutator of (\ref{ga38}) is in the
enveloping algebra as well, the
commutator of two enveloping algebra-valued gauge transformations
remains enveloping algebra-valued. However,  $\mathcal{A}_T$ has an
infinite number of components, expanding order by order. This means
that an enveloping algebra-valued gauge theory would have an infinite
number of degrees of freedom.
This infinite number can be reduced demanding
that the expansion coefficients of enveloping algebra-valued
quantities at every order
depend only on the Lie algebra-valued quantities, i.e. the zeroth order. Then the  NC gauge
theory is determined entirely by the
gauge theory of the commutative space, with the same number of degrees of
freedom. 

We construct the enveloping algebra gauge theory in an expanded way. To perform
the construction we use as a starting point the consistency condition that two consecutive gauge transformations
have to close into another one (\ref{ga32a}). We find that the commutative gauge potential $A^0_\mu$ appears
in the expansion of all quantities of NC gauge theory:
\begin{equation}
\label{ga41aa}
\La_\al:=\La[\al,A^0_\mu], \qquad \psi:=\psi[\psi^0,A^0_\mu ], \qquad A_\mu:=A_\mu[A_\mu^0],\quad \textrm{and}\quad F_{\mu\nu}:=
F_{\mu\nu}[A^0_\mu],\end{equation}
where the square brackets denote
functional dependence. In particular, these quantities depend on the Lie
algebra quantities and an arbitrary number of derivatives on
them. Still, the functionals are supposed to be local in the sense
that at any finite order in the expansion, only a finite number of
derivatives appears.
To avoid notational clutter, we will keep this functional
dependence implicit.

In this
constructive approach, there is  no principle to ensure that the reduction
to the commutative degrees of freedom is indeed possible.
We have to perform an explicit construction order by order in the
parameter of the noncommutativity.

Since $\La_\al$  depends on Lie algebra quantities only,
$\de_{\La_\al}$ reduces to $\de_{\La_\al}=\de_\al$. Since $\La_\al$  depends on  $A^0_\mu$ explicitly, a gauge
transformation of $\La_\al$ is not zero
$\de_{\al}\La_\be\ne0$. Therefore  the consistency relation
(\ref{ga32a}) for 
enveloping algebra-valued 
gauge theory, which must be true for
all fields $\psi$, reads \cite{jmssw}:
\begin{equation}
i\de_{\al} \La_\be -i\de_{\be}\La_\al
+\lb \La_\al\ds\La_\be\rb=i\La_{-i[\al,\be]}.\label{ga43}
\end{equation}
This consistency condition  has the
virtue of being an equation of the gauge parameter $\La_\al$ alone (in
contrast to \cite{sw}).
Once we find solutions for (\ref{ga43}), it is possible to
solve
\begin{equation}
\label{ga44}
\de_\al \psi=i\La_\al\star\psi,\quad \textrm{and}\quad 
\de_\al A_\mu = \pat_\mu \La_\al
-i\lb A^\mu\ds \La_\al\rb, \quad \textrm{etc.}
\end{equation}

Since the enveloping algebra
was introduced because of  the NC $\star$-product,
 we expand
$\La_\al$ in terms of the parameter of the
noncommutativity  $\hbar$:
\begin{equation}
\label{ga45}
\La_\al=\al+\hbar \La_\al^1 +\hbar^2\La_\al^2+\dots\quad.
\end{equation}
Correspondingly,  $\psi$,
$A_\mu$ and $F_{\mu\nu}$ are expanded in terms of $\hbar$
\begin{eqnarray}
\label{ga46}
\psi&=&\psi^0+\hbar \psi^1 +\hbar^2\psi^2+\dots,\\
A_\mu&=&A^0_\mu+\hbar A^1_\mu+\hbar^2A^2_\mu+\dots,   \quad \textrm{etc.}\label{ga46a}
\end{eqnarray}
In particular, we do not expand in the basis of the enveloping algebra.  There
are other suitable expansions, e.g. in terms of the number of factors of gauge
potentials $A^0_\mu$. This expansion allows some interesting
all-orders summations  \cite{jmssw}, \cite{okuyama2}. 

%
%
%
%
%
%
%
%
%
%
%
%
%

\section{$\theta^{\mu\nu}$-expanded fields up to second order}
\label{solutions}
Starting from (\ref{ga43}), we expand the
$\star$-product 
\begin{equation}
\label{ga49}f(x)\star
g(x)=\exp(\frac{i\hbar}{2}\theta^{\mu\nu}\frac{\p}{\p
  y^\mu}\frac{\p}{\p z^\nu})f(y)g(z)|_{y,z\rightarrow x}
\end{equation}
 order by order in $\hbar$, solve the resulting expanded version of
 (\ref{ga43}) at a given order, reinsert the solution
for calculating the next order  etc.  The expansions (\ref{ga45}), (\ref{ga46})
and (\ref{ga46a})
are defined in  such a way that  (\ref{ga44}) coincides in zeroth order with 
(\ref{ga32}) and  (\ref{ga34}). 

We expand (\ref{ga43}) to first order in $\hbar$
\begin{equation}
\label{ga50}
i(\de_\al\La^1_\be
-\de_\be\La^1_\al)+\lb\al,\La^1_\be\rb+
\lb\La^1_\al,\be\rb+ \lb\al\ds\be\rb |_{{\mathcal O}(\hbar^1)}
=i\La^1_{-i[\al,\be]},
\end{equation}
and  insert the explicit form of the Moyal-Weyl $\star$-product to obtain
\begin{equation}
\label{ga51}
\Delta \La^1=i(\de_\al\La^1_\be
-\de_\be\La^1_\al)+\lb\al,\La^1_\be\rb + \lb\La^1_\al,\be\rb - i\La^1_{-i[\al,\be]}=-\frac{i\hbar}{2}\theta^{\mu\nu} \{\partial_\mu \al, \partial_\nu
\be\}  .
\end{equation}
This is an
inhomogeneous linear equation in $\Lambda^1_\al$ with the solution  \cite{jmssw}:
\begin{equation}
\label{ga52}
\La^1_{\al}=-\frac{1}{4}\theta^{\mu\nu} \{A^0_\mu, \partial_\nu
\al\}.
\end{equation}
This solution is not unique, it is always possible to add solutions of
the homogeneous equation $\Delta\La^1=0$,  we defer the discussion of such ambguities
to section \ref{freedom}.

We have introduced $\Delta \La^k=i(\de_\al\La^k_\be
-\de_\be\La^k_\al)+\lb\al,\La^k_\be\rb + \lb\La^k_\al,\be\rb -
i\La^k_{-i[\al,\be]}$  as a shorthand (cp. \cite{zumino1})  in the inhomogeneous equation
 (\ref{ga51}). The
 structure of $\Delta \La^k$ is identical at every 
 order $k$ in $\hbar$.
 
The second order of (\ref{ga43}) reads:
\begin{equation}
\label{ga55}
\Delta \La^2 =\frac{1}{8}\theta^{\mu\nu}\theta^{\ka\la}[\partial_\mu\partial_\ka\al,\partial_\nu\partial_\la\be]-
[\La_\al^1,\La_\be^1]
-\frac{i}{2}\theta^{\mu\nu}\Big(\{\partial_\mu\La^1_{\al},\partial_\nu\be\}
-\{\partial_\mu\La^1_{\be},\partial_\nu\al\}\Big).
\end{equation}
Using (\ref{ga52}) for $\La^1_\al$, we  find the following solution for (\ref{ga55}):
\begin{eqnarray}
\label{ga57}
\La_\al^{2} &=&
\frac{1}{32}\theta^{\mu\nu}
\theta^{\ka\la}\Big(\{A^{0}_{\mu},\{\partial_{\nu}A^{0}_{\ka},\partial_{\la}\al\}\}+\{A^{0}_{\mu},\{A^{0}_{\ka},\partial_{\nu}\partial_{\la}\al\}\}         \\
&&\quad +\{\{A^{0}_{\mu},\partial_{\nu}A^{0}_{\ka}\},\partial_{\la}\al\}\}-\{\{F^0_{\mu\ka},A^{0}_{\nu}\},\partial_{\la}\al\}
-2i[
\partial_{\mu}A^{0}_{\ka},\partial_{\nu}\partial_{\la}\al] \Big).\nonumber
\end{eqnarray}
This solution differs from the one stated in
\cite{jmssw} by a solution of the homogeneous solution at second
order. Similar solutions had been found  in \cite{goto} and \cite{asakawa}.

Next we determine 
 the field $\psi$  (\ref{ga46}) from (\ref{ga44}) up to second order.
To first order in $\hbar$ we obtain:
\begin{equation}
\label{ga58}
\Delta_\al \psi^1:=\de_\al\psi^1-i\al\psi^1=
i\La_\al^1\psi^0-\frac{1}{2}\theta^{\mu\nu}\partial_\mu\al\partial_\nu\psi^0.
\end{equation}
Note the definition of the operator $\Delta_\al  \psi^k$ for any order
$k$: $\Delta_\al\psi^k=\de_\al\psi^k-i\al\psi^k$.
Using (\ref{ga52}) we find 
\begin{equation}
\label{ga59}
\psi^1=-\frac{1}{2}\theta^{\mu\nu}A^0_\mu\partial_\nu\psi^0+\frac{i}{4}\theta^{\mu\nu}A^0_\mu
A^0_\nu\psi^0.
\end{equation}

Similarly the next order, 
\begin{equation}
\label{ga59a}
\Delta_\al \psi^2=
i\La_\al^2\psi^0+i\La_\al^1\psi^1-\frac{1}{2}\theta^{\mu\nu}\partial_\mu\La_\al^1\partial_\nu\psi^0-\frac{1}{2}\theta^{\mu\nu}\partial_\mu\al\partial_\nu\psi^1-\frac{i}{8}\theta^{\mu\nu}\theta^{\ka\la}\partial_\mu\partial_\ka\al\partial_\nu\partial_\la\psi^0,
\end{equation}
is solved by:
\begin{eqnarray}
\label{ga60a}
\psi^{2}&=& -\frac{i}{8} \theta^{\mu\nu}\theta^{\ka\la} \Big(\partial_\ka
A^0_\mu\partial_\nu\partial_\la \psi^0 +iA^0_\ka
A^0_\mu\partial_\nu\partial_\la \psi^0-i\partial_\ka
A^0_\mu A^0_\nu \partial_\la \psi^0 +iF^0_{\ka \mu} A^0_\nu \partial_\la \psi^0
\nonumber\\
&&\qquad-iA^0_\nu\partial_\ka
A^0_\mu  \partial_\la \psi^0+2i A^0_\nu F^0_{\ka \mu} \partial_\la \psi^0+2A^0_\mu A^0_\ka A^0_\nu \partial_\la \psi^0 - A^0_\mu A^0_\nu
A^0_\ka\partial_\la \psi^0\Big)\nonumber\\
&&-\frac{1}{32}\theta^{\mu\nu}\theta^{\ka\la} \Big(2\partial_\ka
A^0_\mu\partial_\la A^0_\nu \psi^0 -2i\partial_\ka
A^0_\mu A^0_\la A^0_\nu\psi^0+2i A^0_\nu A^0_\la\partial_\ka
A^0_\mu\psi^0 \\
&&\quad   +i[[\partial_\ka
A^0_\mu, A^0_\nu], A^0_\la] \psi^0+4i A^0_\nu F^0_{\ka \mu}A^0_\la\psi^0- A^0_\ka A^0_\la A^0_\mu A^0_\nu\psi^0+2 A^0_\ka A^0_\mu A^0_\nu A^0_\la\psi^0\Big).\nonumber
\end{eqnarray}
Note that since we use a $\La_\al^2$ different from the one stated in
\cite{jmssw}, (\ref{ga60a}) differs from the one stated there.
The conjugate field $\overline{\psi}=\psi^\dagger \ga^0$ is obtained by conjugation of
$\psi$, $\overline{\psi}^0=\overline{\psi^0} $ and
$\overline{\psi}^1=\overline{\psi^1}$.

The enveloping
algebra gauge potential (\ref{ga46a}) is determined by expanding
(\ref{ga44}), using $\Delta_\al A^k_\si=\de_\al
A^k_\si-i[\al,A^k_\si]$:
\begin{eqnarray}
\label{ga63}
\Delta_\al A^1_\si&=&\partial_\si\La_\al^1-i[A^0_\si,\La_\al^1]+\frac{1}{2}\theta^{\mu\nu}\{\pat_\mu
A^0_\si, \pat_\nu\al\},\nonumber\\
\Delta_\al
A^2_\si&=&\partial_\si\La_\al^2-i[A^0_\si,\La_\al^2]-i[A^1_\si,\La_\al^1]+\frac{1}{2}\theta^{\mu\nu}\{\pat_\mu
A^1_\si, \pat_\nu\al\}\\
&&+\frac{1}{2}\theta^{\mu\nu}\{\pat_\mu
A^0_\si,
\pat_\nu\La_\al^1\}+\frac{i}{8}\theta^{\mu\nu}\theta^{\ka\la}[\pat_\ka\pat_\mu
A^0_\si,\pat_\la\pat_\nu\al].\nonumber
\end{eqnarray}
A solution for $A^1_\si$ is:
\begin{equation}
\label{ga65}
A^1_\si=-\frac{1}{4}\theta^{\mu\nu}\Big(\{A^0_\mu,\partial_\nu
A^0_\si\}-\{ F^0_{\mu\si},A^0_\nu\}\Big),
\end{equation}
with $F^0_{\mu\nu}$ the Lie algebra field strength (\ref{ga35}).
Using (\ref{ga52}) and (\ref{ga57}), we obtain for $A_{\si}^{2}$:
{\small \begin{eqnarray}
\label{ga67}
A_{\si}^{2}&=&\frac{1}{32}\theta^{\mu\nu}\theta^{\ka\la}\Big(\{\{A^0_\ka,\partial_\la A^0_\mu\},\partial_\nu A^0_\si\}-\{\{F^0_{\ka\mu},
A^0_\la\},\partial_\nu A^0_\si\}-2i[\partial_\ka
A^0_\mu,\partial_\la\partial_\nu A^0_\si]\nonumber\\
&&\quad -\{A^0_\mu, \{\partial_\nu F^0_{\ka \si},
A^0_\la\}\}-\{A^0_\mu, \{ F^0_{\ka \si}, \partial_\nu A^0_\la\}\}  +
\{A^0_\mu, \{\partial_\nu A^0_\ka, \partial_\la A^0_\si\}\}\nonumber\\
&&\quad  + \{A^0_\mu, \{ A^0_\ka, \partial_\nu\partial_\la A^0_\si\}\}
+ \{\{A^0_\ka,\partial_\la F^0_{\mu\si}\}, A^0_\nu\}
-\{\{\mathcal{D}_\ka^0 F^0_{\mu\si}, A^0_\la\}, A^0_\nu\} \\
&&\quad-2\{\{F^0_{\mu\ka},F^0_{\si\la}\}, A^0_\nu\}+2i[\partial_\ka
F^0_{\mu\si},\partial_\la A^0_\nu]-\{F^0_{\mu\si},\{A^0_\ka,
\partial_\la A^0_\nu\}\}+\{F^0_{\mu\si},\{F^0_{\ka
  \nu},A^0_\la\}\}\Big). \nonumber
\end{eqnarray}}
The enveloping
algebra-valued covariant
derivative 
${\cal{D}}_\si\psi = \partial_\si\psi-i A_\si\star \psi$
is obtained immediately, e.g. 
\begin{equation}
\label{ga67fff}
{\cal D}_\si\psi|_{\mathcal{O}(\hbar^1)}=\theta^{\mu\nu}(-\frac{1}{2}A^0_\mu\p_\nu{\cal D}^0_\si\psi^0+\frac{i}{4}A^0_\mu A^0_\nu{\cal D}^0_\si\psi^0+\frac{1}{2}F^0_{\mu\si}{\cal D}^0_\nu\psi^0).
\end{equation}

In order to express  the enveloping
algebra-valued field strength
in terms of Lie algebra-valued quantities, we insert (\ref{ga65}) and (\ref{ga67})
into the definition (\ref{ga40}). We obtain:
\begin{equation}
\label{ga70}
F^1_{\rho\si}=-\frac{1}{4}\theta^{\ka\la}\Big(\{A^0_\ka,\partial_\la
F^0_{\rho\si}\}-\{{\cal D}^0_\ka F^0_{\rho\si}, A^0_\la\}-2\{F^0_{\rho\ka},F^0_{\si\la}\}\Big).
\end{equation}
If we had determined $F^1_{\rho\si}$ from its covariant
transformation behaviour $\de_\al F_{\rho\si}=i\lb\La_\al\ds
F_{\rho\si}\rb$, only the first two terms in
(\ref{ga70}) could have been reproduced. 

In second order we obtain:
\begin{eqnarray}
\label{ga71}
F^2_{\rho\si}&=&\frac{1}{32}\theta^{\mu\nu}\theta^{\ka\la}\Big(2\{\{A^0_\mu,\p_\nu A^0_\ka\},\partial_\la
F^0_{\rho\si}\}-2\{\{F^0_{\mu\ka},
A^0_\nu\},\partial_\la
F^0_{\rho\si}\}\nonumber \\
&&\qquad-2\{A^0_\ka,\partial_\la\{{\cal D}^0_\mu F^0_{\rho\si}, A^0_\nu\}\}-4\{A^0_\ka,\partial_\la\{F^0_{\rho\mu},F^0_{\si\nu}\}\}
 -4i\lb \p_\mu A^0_\ka, \p_\nu\partial_\la
F^0_{\rho\si}\rb \nonumber \\
&&\qquad+i\{\lb\{A^0_\mu,
\p_\nu A^0_\ka\}, F^0_{\rho\si}\rb, A^0_\la\}-i\{\lb\{F^0_{\mu\ka},
A^0_\nu\}, F^0_{\rho\si}\rb, A^0_\la\}\nonumber \\
&&\qquad+i\{\lb A^0_\ka, \{A^0_\mu,\partial_\nu
F^0_{\rho\si}\}\rb, A^0_\la\}-i\{\lb A^0_\ka,\{{\cal D}^0_\mu F^0_{\rho\si}, A^0_\nu\}\rb,
A^0_\la\}\nonumber \\
&&\qquad-2i\{\lb A^0_\ka,\{F^0_{\rho\mu},F^0_{\si\nu}\}\rb, A^0_\la\}+i
\{\lb A^0_\ka,F^0_{\rho\si}\rb,\{A^0_\mu, \p_\nu A^0_\la\}\} \\
&&\qquad-i\{\lb
A^0_\ka,F^0_{\rho\si}\rb,\{F^0_{\mu\la},
A^0_\nu\}\}
+2\{\{\p_\mu A^0_\ka, \p_\nu F^0_{\rho\si}\}, A^0_\la\}\nonumber\\
&&\qquad+2
\lb\p_\mu\lb A^0_\ka,F^0_{\rho\si}\rb, \p_\nu A^0_\la\rb+2\{\{{\cal D}^0_\mu
F^0_{\rho\ka},A^0_\nu\},F^0_{\si\la}\}-2\{\{A^0_\mu,\p_\nu
F^0_{\rho\ka}\},F^0_{\si\la}\}\nonumber\\
&&\qquad+4\{\{F^0_{\mu\rho},F^0_{\nu\ka},\},F^0_{\si\la}\}+2\{F^0_{\rho\ka},\{{\cal D}^0_\mu
F^0_{\si\la},A^0_\nu\}\}-2\{F^0_{\rho\ka},\{A^0_\mu,\p_\nu
F^0_{\si\la}\}\}\nonumber\\
&&\qquad+4\{F^0_{\rho\ka},\{F^0_{\mu\si},F^0_{\nu\la}\}\}+4i \lb\p_\mu F^0_{\rho\ka},\p_\nu F^0_{\si\la}\rb+2\{A^0_\ka,\partial_\la\{ A^0_\mu,\partial_\nu
F^0_{\rho\si}\}\}\Big).\nn
\end{eqnarray}

%
%
%
%
%
%
%
%
%
%
%
%
%

\section{Constructing actions}
\label{build}

We focus on constructing NC
generalisations of the following
Lagrangians\footnote{We use the unusual convention that the trace over
  matrix indices for Yang-Mills is part of the integral, not of the Lagrangian.}: 
\begin{eqnarray}
\label{ac1}
\textrm{Yang-Mills} &&  \hspace{1mm}\mathcal{L}^0_{\textrm{\tiny YM}}=
F^0_{\rho\si}F^{0\rho\si}\\
\textrm{Minimally coupled fermions} &&\mathcal{L}^0_{\textrm{\tiny
    MCF}}= i \overline{\psi^0} \ga^\rho \mathcal{D}^0_\rho \psi^0
\label{ac1a}\\
\textrm{Mass term for fermions} &&  \hspace{1mm}\mathcal{L}^0_{\textrm{\tiny MF}}=m\overline{\psi^0}\psi^0\label{ac1b}.
\end{eqnarray}
Although a fermionic mass term is not part of the Standard Model, it
suffices to
illustrate the properties of NC gauge theory. The remaining Standard Model
Lagrangians of a scalar mass term, a scalar quartic potential,
minimally coupled kinetic terms for the scalar field and the Yukawa
potential can be obtained almost immediately
using the results of the above mentioned three cases. The only
additional complication is the so-called hybrid Seiberg-Witten map
(cp. \cite{cjssw}).

Requiring that the NC generalisations of (\ref{ac1}) to (\ref{ac1b})
are the
Standard Model Lagrangians in zeroth order in $\hbar$, the most natural approach is to replace
every field by its NC  analogue. Then the NC Lagrangians can be
written using the $\star$-product and the
expansions of the enveloping algebra-valued fields:
\begin{eqnarray}
\label{ac2}
 \hspace{1mm}\mathcal{L}_{\textrm{\tiny YM}}&=&
F_{\rho\si}\star F^{\rho\si}, \\
\mathcal{L}_{\textrm{\tiny
    MCF}}&=&i\overline{\psi} \star \ga^\rho \mathcal{D}_\rho \psi,
\label{ac2a}\\
\mathcal{L}_{\textrm{\tiny
    MF}}&=&m\overline{\psi}\star \psi.
\label{ac2b}
\end{eqnarray}

First we discuss mass terms and gauge coupling terms for fermion fields $\psi$ transforming from the
left. The zeroth orders
coincide with the commutative Lagrangians by definition, the first and
second order read:
\begin{eqnarray}
\label{ac6}
m\overline{\psi}\star\psi|_{\mathcal{O}(\hbar^1)}&=&
m\left(\overline{\psi^0}\star^1\psi^0+ \overline{\psi^1}\psi^0+
\overline{\psi^0}\psi^1\right)=
m \frac{i}{2}\theta^{\mu\nu}{\cal{D}}_\mu^0\overline{\psi}^0{\cal{D}}^0_\nu\psi^0,
\\
m\overline{\psi}\star\psi|_{\mathcal{O}(\hbar^2)}&=&
m\left(\overline{\psi^0}\star^2\psi^0+ \overline{\psi^1}\star^1\psi^0+\overline{\psi^0}\star^1\psi^1+\overline{\psi^1}\psi^1+\overline{\psi^2}\psi^0+\overline{\psi^0}\psi^2\right)=\nonumber
\\&=&m\left(-\frac{1}{8}\theta^{\mu\nu}\theta^{\ka\la}{\cal{D}}^0_\mu{\cal{D}}^0_\ka\overline{\psi^0}{\cal{D}}^0_\nu{\cal{D}}^0_\la\psi^0-\frac{i}{4}\theta^{\mu\nu}\theta^{\ka\la}{\cal{D}}^0_\mu\overline{\psi^0}F^0_{\nu\ka}{\cal{D}}^0_\la\psi^0\right).
\end{eqnarray}
In these expressions the Lie algebra-valued covariant derivative is
evaluated on a conjugate
field as
${\cal{D}}^0_\mu\overline{\psi^0}=\p_\mu\overline{\psi^0}+i\overline{\psi^0}A^0_\mu$.

The minimal coupling of matter fields to
 gauge potentials reads:
{\small \begin{eqnarray}
\label{ac7}
i\overline{\psi}\star\gamma^\rho{\cal{D}}_\rho\psi|_{\mathcal{O}(\hbar^1)}&=&-\frac{1}{2}\theta^{\mu\nu}{\cal{D}}^0_\mu\overline{\psi^0}\gamma^\rho{\cal{D}}^0_\nu{\cal{D}}^0_\rho\psi^0
+\frac{i}{2}\theta^{\mu\nu}\overline{\psi^0}\gamma^\rho F^0_{\mu\rho}{\cal{D}}^0_\nu\psi^0, \\
i\overline{\psi}\star\gamma^\rho{\cal{D}}_\rho\psi|_{\mathcal{O}(\hbar^2)}&=&-\frac{i}{8}\theta^{\mu\nu}\theta^{\ka\la}{\cal{D}}^0_\ka{\cal{D}}^0_\mu\overline{\psi^0}\gamma^\rho
{\cal{D}}^0_\la{\cal{D}}^0_\nu{\cal{D}}^0_\rho\psi^0
+\frac{1}{4}\theta^{\mu\nu}\theta^{\ka\la}{\cal{D}}^0_\ka\overline{\psi^0}F^0_{\la\mu}{\cal{D}}^0_\nu{\cal{D}}^0_\rho\psi^0
\nonumber\\
&&-\frac{1}{4}\theta^{\mu\nu}\theta^{\ka\la}{\cal{D}}^0_\ka\overline{\psi^0}\gamma^\rho{\cal{D}}^0_\la\left(F^0_{\mu\rho}{\cal{D}}^0_\nu\psi^0\right)-\frac{1}{8}\theta^{\mu\nu}\theta^{\ka\la}\overline{\psi^0}\gamma^\rho
\left({\cal{D}}^0_\ka F^0_{\mu\rho}\right){\cal{D}}^0_\la{\cal{D}}^0_\nu\psi^0
\nonumber\\
&&-\frac{i}{8}\theta^{\mu\nu}\theta^{\ka\la}\overline{\psi^0}\gamma^\rho
F^0_{\mu\ka}F^0_{\la\rho}{\cal{D}}^0_\nu\psi^0-\frac{i}{4}\theta^{\mu\nu}\theta^{\ka\la}\overline{\psi^0}\gamma^\rho
F^0_{\mu\rho}F^0_{\nu\ka}{\cal{D}}^0_\la\psi^0.
\end{eqnarray}   }

Next we consider the NC Yang-Mills Lagrangian. The zeroth order again is by
definition identical to its
commutative counterpart. For higher orders we only need the
$\theta^{\mu\nu}$-expansion at $j$th order
$F_{\rho\si}^j$ with lower indices, for the dual field strength indices $F^{j\rho\si}$  can be raised
 with a formal metric (we assume a flat NC space):
\begin{equation}
F_{\rho\si} \star F^{\rho\si}= g^{\al\rho}g^{\be\si} F_{\al\be} \star F_{\rho\si} ,  
\end{equation}
therefore the order $F_{\rho\si}\star F^{\rho\si}$
vs. $F^{\rho\si}\star F_{\rho\si}$ is unimportant. In first order in
$\hbar$ we obtain
\begin{eqnarray}
\label{ac3}
F_{\rho\si}\star
F^{\rho\si}|_{\mathcal{O}(\hbar^1)}&=&\frac{i}{2}\theta^{\mu\nu}{\cal{D}}^0_\mu F_{\rho\si}^0{\cal{D}}^0_\nu F^{0\rho\si}+\frac{1}{2}\theta^{\mu\nu}\{\{F_{\rho\mu}^0,F_{\si\nu}^0\},F^{0\rho\si}\}  \\
&&-\frac{1}{2}\theta^{\mu\nu}\{A^0_\mu,\p_\nu(F_{\rho\si}^0F^{0\rho\si})\}+\frac{i}{4}\theta^{\mu\nu}\{A^0_\mu,\lb
A^0_\nu,\left(F_{\rho\si}^0F^{0\rho\si}\right)\rb\}.\nn
\end{eqnarray}
Similarly, we obtain in second order:
\begin{eqnarray}
\label{ac4}
&&F_{\rho\si}\star
F^{\rho\si}|_{\mathcal{O}(\hbar^2)}= \theta^{\mu\nu}\theta^{\ka\la}\times\nn\\
&&\qquad\Big(-\frac{1}{8}{\cal{D}}^0_\mu{\cal{D}}^0_\ka
F_{\rho\si}^0{\cal{D}}^0_\nu{\cal{D}}^0_\la
F^{0\rho\si}+\frac{i}{4} \lb{\cal{D}}^0_\ka\{F_{\rho\mu}^0,F_{\si\nu}^0\},{\cal{D}}^0_\la
F^{0\rho\si}\rb  \nonumber\\
&&\qquad+\frac{i}{8} \{\lb{\cal{D}}^0_\ka F_{\rho\mu}^0,{\cal{D}}^0_\la
F_{\si\nu}^0\rb,
F^{0\rho\si}\} +\frac{1}{4} \{\{\{ F_{\mu\ka}^0,F^{0}_{\nu\rho}\},F^0_{\la\si}\},F^{0\rho\si}\} \nonumber\\
&&\qquad-\frac{1}{4} \{A^0_\ka,\p_\la\{\{F_{\rho\mu}^0,F_{\si\nu}^0\},F^{0\rho\si}\}\}+\frac{i}{8} \{A^0_\ka,\lb
A^0_\la,\{\{F_{\rho\mu}^0,F_{\si\nu}^0\},F^{0\rho\si}\}\rb\}\nonumber\\
&&\qquad-\frac{i}{4} \{A^0_\ka,\p_\la\left({\cal{D}}^0_\mu
  F_{\rho\si}^0{\cal{D}}^0_\nu
F^{0\rho\si}\right)\}-\frac{1}{8} \{A^0_\ka,\lb
A^0_\la,\left({\cal{D}}^0_\mu
  F_{\rho\si}^0{\cal{D}}^0_\nu
F^{0\rho\si}\right)\rb\}\nonumber\\
&&\qquad+\frac{1}{8} \{A^0_\ka,\p_\la\{A^0_\mu,\p_\nu\left(F_{\rho\si}^0F^{0\rho\si}\right)\}\}-\frac{i}{16} \{A^0_\ka,\lb
A^0_\la,\{A^0_\mu,\p_\nu\left(F_{\rho\si}^0F^{0\rho\si}\right)\}\rb\}\nonumber\\
&&\qquad-\frac{i}{16} \{A^0_\ka,\p_\la\{A^0_\mu,\lb  A^0_\nu,\left(F_{\rho\si}^0F^{0\rho\si}\right)\rb\}\}-\frac{1}{32} \{A^0_\ka,\lb
A^0_\la,\{A^0_\mu,\lb
A^0_\nu,\left(F_{\rho\si}^0F^{0\rho\si}\right)\rb\}\rb\}\nonumber\\
&&\qquad+\frac{1}{8} \{\{A^0_\ka,\p_\la A^0_\mu\},\p_\nu  \left(F_{\rho\si}^0F^{0\rho\si}\right)\}-\frac{1}{16} \{\{A^0_\ka,\p_\mu A^0_\la\},\p_\nu  \left(F_{\rho\si}^0F^{0\rho\si}\right)\}\\
&&\qquad-\frac{i}{16} \{\{A^0_\ka,\p_\la
A^0_\mu\},\lb A^0_\nu,\left(F_{\rho\si}^0F^{0\rho\si}\right)\rb\}+\frac{i}{32} \{\{A^0_\ka,\p_\mu A^0_\la\},\lb A^0_\nu, \left(F_{\rho\si}^0F^{0\rho\si}\right)\rb\}\nonumber\\
&&\qquad-\frac{i}{16} \{A^0_\ka,\lb\{A^0_\mu,\p_\nu A^0_\la\},\left(F_{\rho\si}^0F^{0\rho\si}\right)\rb\}+\frac{i}{32} \{A^0_\ka,\lb\{A^0_\mu,\p_\la A^0_\nu\},\left(F_{\rho\si}^0F^{0\rho\si}\right)\rb\}\nonumber\\
&&\qquad-\frac{1}{32} \{A^0_\ka,\lb\{A^0_\mu,\lb
A^0_\nu,A^0_\la\rb\},\left(F_{\rho\si}^0F^{0\rho\si}\right)\rb\}-\frac{1}{32} \{\{A^0_\ka,\lb
A^0_\la,
A^0_\mu\rb \},\lb A^0_\nu, \left(F_{\rho\si}^0F^{0\rho\si}\right)\rb\}\nonumber\\
&&\qquad-\frac{i}{16} \{\{A^0_\ka,\lb
A^0_\la,
A^0_\mu\rb \},\p_\nu
\left(F_{\rho\si}^0F^{0\rho\si}\right)\}+\frac{i}{8} \lb\{{\cal{D}}^0_\ka
  F^0_{\rho\si}, F^0_{\mu\la}\},{\cal{D}}^0_\nu F^{0\rho\si}\rb\nonumber\\
&&\qquad-\frac{i}{8} \lb\p_\ka A^0_\mu,\p_\la \p_\nu
\left(F_{\rho\si}^0F^{0\rho\si}\right)\rb-\frac{1}{16} \lb\p_\ka
A^0_\mu,\p_\la \lb A^0_\nu,
  \left(F^0_{\rho\si}, F^{0\rho\si}\right)\rb\rb\nonumber\\
&&\qquad-\frac{1}{16} \{A_\ka^0,\{\p_\mu
A^0_\la, \p_\nu
\left(F_{\rho\si}^0F^{0\rho\si}\right)\}\}+\frac{1}{8}\{\{ F_{\mu\rho}^0,F^{0}_{\nu\si}\},\{F^{0\hspace{1mm}\rho}_{\ka},F^{0\hspace{1mm}\si}_{\la}\}\}\Big).\nonumber
\end{eqnarray}

The construction of an action  requires the definition of an
integral. Integration is in general difficult to implement for NC
spaces. It is misleading to expect that the integral is related to 
summing the field values over points, since NC
spaces are ``pointless''. Therefore the usual intuitions
about integration have to be dropped; specific desired properties have to be imposed.
In particular,  the integral
should have the
trace property, such that  a
gauge-covariant Lagrangian implies a gauge invariant action. We demand: 
 \begin{equation}
\label{ac8}
\int\textrm{d}x\hspace{2mm} f \star g  = \int\textrm{d}x\hspace{2mm} g \star f .
\end{equation}
If the integral allows the use of Stokes' theorem, we may partially integrate the derivatives of the
$\star$-product for $\theta^{\mu\nu}=const$ and obtain, because of 
antisymmetry and $x$-independence of $\theta^{\mu\nu}$,
 \begin{equation}
\label{ac9}
\int\textrm{d}x\hspace{2mm} f \star g   = \int\textrm{d}x\hspace{2mm} f g = \int \textrm{d}x\hspace{2mm} g f = \int\textrm{d}x\hspace{2mm} g \star f .
\end{equation}
In addition to the integral, we need for the Yang-Mills action a trace over the matrix indices of the generators
of the underlying non-Abelian gauge group:
 \begin{equation}
\label{ac10a}
\mathcal{S}_{\textrm{\tiny YM}}=\tilde{c}\hspace{1mm}\textrm{Tr}\int\textrm{d}x\hspace{1mm}F_{\rho\si}
\star F^{\rho\si}.
\end{equation}
Altogether this ensures
gauge invariance of the Yang-Mills action:
\begin{equation}
\label{ac11}
\de_\al \mathcal{S}_{\textrm{\tiny YM}}=
\tilde{c}\hspace{1mm}\textrm{Tr}\int\textrm{d}x\hspace{1mm} \Big(i\lb
\La_\al\ds F_{\rho\si}\rb
\star F^{\rho\si}+i F_{\rho\si}\star \lb
\La_\al\ds F^{\rho\si}\rb \Big)=0,\nn
\end{equation}
Since the sum over  generators is not performed, the numerical factor
of $\mathcal{S}_{\textrm{\tiny YM}}$ in (\ref{ac10a}) is an arbitrary real constant
$\tilde{c}$. In the NC enveloping algebra gauge theory,
the choice of
trace for the NC Yang-Mills action is restricted only by
$\frac{1}{g^2}=\sum_\rho c_\rho \textrm{Tr}(\rho(T^a)\rho(T^a))$
(cp. \cite{cjssw}, \cite{aschieri}).
Here $g$ is the coupling of the gauge group, $\rho$ denotes a representation of the generators of the Lie
algebra and the parameters $c_\rho$ are restricted only by the mentioned
condition.

Obviously (\ref{ac3}) and (\ref{ac4}) are not explicitly
gauge covariant. This lack of covariance is cured in the action. Under
an integral the non-covariant terms are rearranged into a field
strength:
\begin{equation}
\label{ac5}
\textrm{Tr}\int\textrm{d}x\hspace{1mm} \theta^{\ka\la}\left(\{A^0_\ka, \p_\la
X\}-\frac{i}{2}\{A^0_\ka,\lb A^0_\la,
X\rb\}\right)=\textrm{Tr}\int\textrm{d}x\hspace{1mm}
\theta^{\ka\la}F^0_{\ka\la} X.
\end{equation}

Partial integration and the trace
property lead to the following  result for the $\hbar$-expanded Yang-Mills Lagrangian:
\begin{eqnarray}
\label{ac12}
&&\tilde{c}\hspace{1mm} \textrm{Tr}\int\textrm{d}x\hspace{1mm} F_{\rho\si} \star
 F^{\rho\si}|_{{\cal O}(\hbar^1)}=\tilde{c}\theta^{\mu\nu}\textrm{Tr}\int\textrm{d}x\hspace{1mm} \big(2 F^0_{\rho\mu}F^0_{\si\nu} F^{0 \rho\si}-\frac{1}{2}F^0_{\mu\nu}F^0_{\rho\si} F^{0 \rho\si} \big),\\
&&\tilde{c}\hspace{1mm} \textrm{Tr}\int\textrm{d}x\hspace{1mm} F_{\rho\si} \star
 F^{\rho\si}|_{{\cal
 O}(\hbar^2)}=\nn\\&&\qquad\tilde{c}\theta^{\mu\nu}\theta^{\ka\la}\textrm{Tr}\int\textrm{d}x\hspace{1mm}\Big(\frac{1}{8}
 F^0_{\mu\nu}F^0_{\ka\la}F^0_{\rho\si} F^{0 \rho\si}-\frac{i}{4}
 F^0_{\mu\nu}(\mathcal{D}^0_\ka F^0_{\rho\si})(\mathcal{D}^0_\la  F^{0
 \rho\si})\nn\\&&\qquad\qquad\qquad\quad\qquad- \frac{1}{8}(\mathcal{D}^0_\mu\mathcal{D}^0_\ka F^0_{\rho\si})
 (\mathcal{D}^0_\nu\mathcal{D}^0_\la F^{0 \rho\si})+
 \frac{i}{2}(\mathcal{D}^0_\mu F^0_{\rho\ka})(\mathcal{D}^0_\nu F^0_{\si\la})F^{0 \rho\si}\nn\\&&\qquad\qquad\qquad\quad\qquad+\frac{1}{2}
 F^0_{\mu\rho}F^0_{\nu\si}F^{0\hspace{1mm}\rho}_{\ka}F^{0\hspace{1mm}\si}_{\la}+\frac{1}{2}
 F^0_{\mu\rho}F^0_{\nu\si}F^{0\hspace{1mm}\si}_{\ka}F^{0\hspace{1mm}\rho}_{\la}\\&&\qquad\qquad\qquad\quad\qquad-\frac{1}{2}
 F^0_{\mu\nu}F^0_{\ka\rho}F^0_{\la\si} F^{0 \rho\si}-\frac{1}{2}
F^0_{\ka\rho}F^0_{\la\si} F^0_{\mu\nu} F^{0 \rho\si}\nn\\&&\qquad\qquad\qquad\quad\qquad+\frac{1}{2}(F^0_{\mu\ka}F^0_{\nu\rho}F^0_{\la\si}+2F^0_{\nu\rho}F^0_{\mu\ka}F^0_{\la\si}+F^0_{\la\si}F^0_{\nu\rho}F^0_{\mu\ka})F^{0 \rho\si}\Big).\nn
\end{eqnarray}

The fermionic mass term reads:
\begin{eqnarray}
\label{ac13}
m\int\textrm{d}x\hspace{1mm} \overline{\psi}\star\psi|_{{\cal O}(\hbar^1)}&=& -\frac{m}{4}\theta^{\mu\nu}\int\textrm{d}x\hspace{1mm}
 \overline{\psi^0} F^0_{\mu\nu}\psi^0,\\
m\int\textrm{d}x\hspace{1mm} \overline{\psi}\star\psi|_{{\cal O}(\hbar^2)}&=& m\theta^{\mu\nu}\theta^{\ka\la}\int\textrm{d}x\hspace{1mm}\big(
\frac{i}{8} \overline{\psi^0} (\mathcal{D}^0_\ka F^0_{\la\mu})\mathcal{D}^0_\nu\psi^0-\frac{1}{8} \overline{\psi^0} F^0_{\ka\mu} F^0_{\la\nu}\psi^0\nonumber\\&&\qquad\qquad\qquad+\frac{1}{32} \overline{\psi^0}  F^0_{\ka\la} F^0_{\mu\nu}\psi^0\big),\label{ac13a}
\end{eqnarray}
and the minimally gauge coupled fermionic action is:
\begin{eqnarray}
\label{ac14}
i\int\textrm{d}x\hspace{1mm} \overline{\psi}\star\gamma^\rho
 {\cal{D}}_\rho\psi|_{{\cal O}(\hbar^1)}&=& \theta^{\mu\nu}\int\textrm{d}x\hspace{1mm}\Big(-\frac{i}{4}
 \overline{\psi}^0\gamma^\rho F^0_{\mu\nu} {\cal{D}}^0_\rho\psi^0
-\frac{i}{2}
 \overline{\psi^0}\gamma^\rho F_{\rho\mu}^0 {\cal{D}}^0_\nu\psi^0\Big),\\
i\int\textrm{d}x\hspace{1mm} \overline{\psi}\star\gamma^\rho
 {\cal{D}}_\rho\psi|_{{\cal O}(\hbar^2)}&=& \theta^{\mu\nu} \theta^{\ka\la}\int\textrm{d}x\hspace{1mm}
 \Big(-\frac{i}{8}\overline{\psi^0}\gamma^\rho
F^0_{\mu\ka}F^0_{\la\rho}{\cal{D}}^0_\nu\psi^0-\frac{i}{4}\overline{\psi^0}\gamma^\rho
F^0_{\mu\rho}F^0_{\nu\ka}{\cal{D}}^0_\la\psi^0\nn\\
&&\qquad\qquad-\frac{1}{8}\overline{\psi^0}\gamma^\rho({\cal{D}}^0_\mu F^0_{\nu\ka})
 {\cal{D}}^0_\la{\cal{D}}^0_\rho\psi^0-\frac{i}{8}\overline{\psi^0}\gamma^\rho
 F^0_{\ka\mu}  F^0_{\la\nu}{\cal{D}}^0_\rho\psi^0\nn\\
&&\qquad\qquad-\frac{1}{4}\overline{\psi^0}\gamma^\rho({\cal{D}}^0_\mu F^0_{\ka\rho}
) {\cal{D}}^0_\nu{\cal{D}}^0_\la\psi^0-\frac{i}{8}\overline{\psi^0}\gamma^\rho
 F^0_{\mu\nu}  F^0_{\ka\rho}{\cal{D}}^0_\la\psi^0\nn\\
&&\qquad\qquad+\frac{i}{32}\overline{\psi^0}\gamma^\rho
 F^0_{\ka\la}  F^0_{\mu\nu}{\cal{D}}^0_\rho\psi^0\Big).\label{ac14a}\end{eqnarray}

%
%
%
%
%

\section{Ambiguities of enveloping algebra gauge theory}
\label{freedom}
We stated after equation (\ref{ga52}) that there is an ambiguity in the construction of
the 
enveloping algebra-valued gauge parameter $\La_\al$. 
In addition, there are
ambiguities in  constructing
fields and the gauge potential. In this section we investigate these
ambiguities in first and second order.  

The ambiguity (\ref{ga52})
has been discussed shortly in \cite{jmssw}, along with a discussion of 
 field redefinitions. Ambiguities in the 
 Seiberg-Witten map have been discussed in \cite{goto},
 \cite{asakawa2}, \cite{brandt2} and
 \cite{zumino3}. However, these approaches have not discussed the
 meaning of the ambiguities for the construction of actions (the
 Yang-Mills action was discussed in
 \cite{goto}).

All ambiguities have
to be solutions of  homogeneous equations:
\begin{eqnarray}
\label{fr1}
\Delta \La^n&=&0,\nn\\\Delta_\al \psi^n&=&0,\\
\Delta_\al A^n_\si&=&0.\qquad\nn
\end{eqnarray}
There are essentially
two types of ambiguities, which have been called covariant and gauge
ambiguities, respectively. However, we do not want to cover
the ambiguities encyclopedically in this note, rather
discuss only those relevant for the actions.

We therefore demand the following requirements for the ambiguities
relevant here. They should be
\begin{itemize}
\item {\it hermitian}, e.g. $\overline{\La}_\al=\La_\al$.
\item {\it of the same index structure as the actions}: That means
  that we do not allow additional factors of a metric to lower the
  indices in $\theta^{\mu\nu}$, e.g. as in \cite{bichl2}.
\item {\it not derivative-valued}: Such terms, e.g. 
    $\La_\al^1=i\theta^{\mu\nu}\p_\mu\al \p_\nu$, are hermitian and
    can be used to consistently solve also higher-order terms (e.g. 
    $\La_\al^2$). However, they should be discussed in a different context.
\end{itemize}

The only first-order hermitian ambiguity for $\La_\al^1$  is therefore:
\begin{equation}
\label{fr5}
\La_\al^{1,c_1}=ic_1\theta^{\mu\nu}\lb A^0_\mu, \p_\nu \al\rb, \quad
c_1 \in \mathbb{R}.
\end{equation} 
This ambiguity $\La_\al^1$   leads to an additional term for the fermion field
\begin{equation}
\label{fr7}
\psi^{1,c_1}=-c_1\theta^{\mu\nu} A^0_\mu A^0_\nu \psi^0, \qquad
\textrm{and}\qquad \overline{\psi^{1,c_1}}=c_1\theta^{\mu\nu}   \overline{\psi}^0A^0_\mu A^0_\nu,
\end{equation} 
and to an additional term for the gauge
potential:
\begin{equation}
\label{fr8}
A^{1,c_1}_\rho=ic_1\theta^{\mu\nu} \big(\lb\p_\rho A^0_\mu,
A^0_\nu\rb-i\lb\lb A^0_\rho, A^0_\mu\rb, A^0_\nu\rb\big)=ic_1\theta^{\mu\nu} \big(\p_\rho( A^0_\mu
A^0_\nu)-i\lb A^0_\rho, A^0_\mu A^0_\nu\rb\big).
\end{equation} 
Adding up these terms, the additional fermionic Lagrangians
$\mathcal{L}^{1,c_1}_{\textrm{\tiny MF}}$ and 
$\mathcal{L}^{1,c_1}_{\textrm{\tiny MCF}}$ derived from (\ref{fr5}) 
are identically zero.
The field
strength corresponding to (\ref{fr8}) is
\begin{equation}
\label{fr9}
F^{1,c_1}_{\rho\si}=c_1\theta^{\mu\nu} \lb\lb F^0_{\rho\si},
A^0_\mu\rb, A^0_\nu\rb,
\end{equation} 
and the corresponding Yang-Mills action is therefore
\begin{equation}
\label{fr10}
\mathcal{S}^{1,c_1}_{\textrm{\tiny
    YM}}=-c_1\tilde{c}\theta^{\mu\nu}\textrm{Tr}\int \mathrm{d}x\hspace{1mm}(A^0_\mu A^0_\nu
F^0_{\rho\si}F^{0\rho\si}-F^0_{\rho\si}F^{0\rho\si}A^0_\mu
A^0_\nu)=0.
\end{equation} 
Therefore the ambiguity $\La_\al^{1,c_1}$ does not contribute at all to the
action. 

Additional ambiguities at first order are solutions of the homogeneous
equations (\ref{fr1}) for $\psi^1$,
\begin{equation}
\label{fr15}
\psi^{1,c_2}=c_2 \theta^{\mu\nu} F^0_{\mu\nu}\psi^0,
\end{equation} 
and the gauge potential $A_\rho^1$,
\begin{equation}
\label{fr16}
A^{1,c_3}_\rho=c_3 \theta^{\mu\nu} \mathcal{D}^0_\rho F^0_{\mu\nu}.\\
\end{equation} 
That this is the only ambiguity w.r.t. the gauge potential can be seen using the Bianchi identity. 
The field strength corresponding to $A^{1c_3}_\rho$ is
\begin{equation}
\label{fr17}
F^{1,c_3}_{\rho\si}=i\lb\mathcal{D}^{1,c_3}_\rho,\mathcal{D}^{0}_\si\rb+i\lb\mathcal{D}^{0}_\rho,\mathcal{D}^{1,c_3}_\si\rb=-i
c_3 \theta^{\mu\nu}\lb F^0_{\rho\si},F^0_{\mu\nu}\rb. 
\end{equation} 
Since this field strength is a commutator, it does not contribute to
the Yang-Mills action.  There are no additional ambiguities than
(\ref{fr17}) for the field strength, since we demanded that it is strictly calculated
from the gauge potential and not constructed as a solution of the transformation
law. While the Yang-Mills action  at first order is 
unambiguous, 
(\ref{fr15}) and (\ref{fr16}) introduce an ambiguity in the fermionic action:
\begin{equation}
\label{fr18}
\mathcal{S}^{1, c_2, c_3}_{\textrm{\tiny{MCF, MF}}}=\int
\textrm{d}n x\hspace{1mm} \Big( 2c_2
\theta^{\mu\nu}\overline{\psi^0}
F^0_{\mu\nu}(i\gamma^\rho\mathcal{D}^0_\rho -m)\psi^0+ (c_2+c_3)
\theta^{\mu\nu}\overline{\psi^0} i\gamma^\rho(\mathcal{D}^0_\rho
F^0_{\mu\nu}) \psi^0\Big).
\end{equation} 
We add (\ref{fr18}) to the fermionic
action derived in the previous section (\ref{ac13}) and (\ref{ac14}):
\begin{eqnarray}
\label{fr19}
\mathcal{S}^{1}_{\textrm{\tiny{MCF, MF}}}+\mathcal{S}^{1, c_2, c_3}_{\textrm{\tiny{MCF, MF}}} &=&\int
\textrm{d}^n x \Big((\underbrace{2c_2-\frac{1}{4}}_{d_1})
\theta^{\mu\nu}\overline{\psi^0}
F^0_{\mu\nu}(i\gamma^\rho\mathcal{D}^0_\rho -m)\psi^0\\&&\qquad-\frac{i}{2}
\theta^{\mu\nu}\overline{\psi^0}\gamma^\rho F^0_{\rho\mu}\mathcal{D}^0_\nu
 \psi^0+ (\underbrace{ic_2+c_3}_{d_2})\theta^{\mu\nu}\overline{\psi^0} \gamma^\rho(\mathcal{D}^0_\rho F^0_{\mu\nu})\nonumber
 \psi^0 \Big).
\end{eqnarray} 

Choosing $c_2=\frac{1}{8}$ and $c_3=-\frac{i}{8}$ two terms can
be set to zero, but choosing any other value for $d_1$ and $d_2$ will
be just as consistent with the structure of the enveloping algebra gauge
theory. 

Since physics must not depend on a choice of parameters, 
any prediction based on a particular value of $d_i$ is
unphysical. This 
type of ambiguity w.r.t. the value of $c_2$ and $c_3$ or $d_1$ and $d_2$ is called a field redefinition \cite{jmssw}, \cite{zumino3},
\cite{brandt1}. We may conclude that the only
physically relevant 
term in the NC fermionic action at first order in $\hbar$ is 
 \begin{equation}
\label{fr19a}
\mathcal{S}^{1, \textrm{\tiny{relevant}}}_{\textrm{\tiny{MCF, MF}}} =-\frac{i}{2}\theta^{\mu\nu}\int
\textrm{d}^n x 
\overline{\psi^0}\gamma^\rho F^0_{\rho\mu}\mathcal{D}^0_\nu
 \psi^0,
\end{equation} 
since the two other terms are proportional to a field redefinition. In particular,
no mass term appears at first order in $\hbar$. To repeat, all terms
of the NC Yang-Mills
action at first order in $\theta^{\mu\nu}$ are physically relevant.

Next let us turn to the ambiguities appearing at second order. First
we investigate the effect of  first-order ambiguties at second order.
The ambiguity in the gauge parameter $\La_\al^{1,
  c_1}$ leads to the following additional terms in $\La_\al^2$
\begin{eqnarray}
\label{fr24a}
\La_\al^{2, c_1}&=&c_1\theta^{\mu\nu}\theta^{\ka\la}\Big(-\frac{i}{2}\{\lb \pat_\mu A^0_\ka, A^0_\la\rb, \pat_\nu \al\}-\frac{1}{4}\{\lb\lb A^0_\mu, \pat_\ka \al\rb, A^0_\nu\rb, A^0_\la\} \nn\\&&\qquad+\frac{1}{4}\{\lb\lb A^0_\mu, A^0_\ka\rb, A^0_\nu\rb, \pat_\la \al\} \Big)-\frac{c_1^2i}{4}\theta^{\mu\nu}\theta^{\ka\la}\lb \lb A^0_\mu, A^0_\nu \rb, \lb A^0_\ka , \pat_\la \al\rb\rb. 
\end{eqnarray} 
The gauge parameter ambiguities $\La_\al^{1, c_1}$ and 
$\La_\al^{2, c_1}$ lead to additional terms in
$\psi^{2,c_1}$:
\begin{eqnarray}
\label{fr25}
\psi^{2,c_1}&=&c_1\theta^{\mu\nu}\theta^{\ka\la}\Big(-\frac{i}{2}\lb
\pat_\mu A^0_\ka, A^0_\la\rb \pat_\nu\psi^0 +\frac{1}{2} A^0_\mu
A^0_\nu A^0_\ka\pat_\la \psi^0 -\frac{i}{4} A^0_\mu A^0_\nu A^0_\ka
A^0_\la \psi^0\Big)\nn\\
&&+\frac{c_1^2}{2}\theta^{\mu\nu}\theta^{\ka\la} A^0_\mu A^0_\nu A^0_\ka
A^0_\la \psi^0.
\end{eqnarray} 
and the gauge potential 
\begin{eqnarray}
\label{fr25a}
A^{2,c_1}_\rho&=&c_1\theta^{\mu\nu}\theta^{\ka\la}\Big(-\frac{i}{2}\{\lb
\pat_\mu A^0_\ka, A^0_\la\rb \pat_\nu A^0_\rho\} +\frac{1}{4}\{\lb\lb
A^0_\mu, A^0_\ka\rb, A^0_\nu\rb, (F^0_{\la\rho}+\pat_\la A^0_\rho)\}\nn\\
&&\qquad -\frac{1}{4}\{\lb\lb
A^0_\mu,(F^0_{\ka\rho}+\pat_\ka A^0_\rho) \rb, A^0_\nu\rb, A^0_\la\}\Big)
\\
&&-\frac{c_1^2}{4}\theta^{\mu\nu}\theta^{\ka\la}\Big(i\lb\lb A^0_\mu,
A^0_\nu\rb,\lb \pat_\rho A^0_\ka, A^0_\la\rb\rb +\lb\lb A^0_\mu,
A^0_\nu\rb,\lb \lb A^0_\rho, A^0_\ka\rb , A^0_\la\rb\rb \Big).\nn
\end{eqnarray} 
Plugging these terms into the second order fermionic mass and minimally coupled action,
all contributions (in $c_1^2$ and in $c_1$) drop out. Similarly these ambiguities
in $c_1$  do not contribute to the
Yang-Mills action.

We have analysed in addition  three second-order ambiguities in $\La_\al^2$
\begin{eqnarray}
\label{fr26}
\La_\al^{2, c_4}&=& c_4 \theta^{\mu\nu}\theta^{\ka\la}\Big(\{\pat_\mu A^0_\ka, \pat_\nu\pat_\la \al\}+i\{\pat_\mu A^0_\ka, \lb\pat_\nu\al, A^0_\la \rb\}\Big),\nn\\
\La_\al^{2, c_5}&=& c_5 \theta^{\mu\nu}\theta^{\ka\la}\{\pat_\mu
A^0_\nu, \lb A^0_\ka, \pat_\la \al\rb\},  \\
\La_\al^{2, c_6}&=& c_6 \theta^{\mu\nu}\theta^{\ka\la}\{\lb A^0_\mu, A^0_\nu\rb, \lb A^0_\ka, \pat_\la \al\rb\}.\nn
\end{eqnarray} 
These lead to the following terms for fields $\psi$:
\begin{eqnarray}
\label{fr27}
\psi^{2, c_4}&=& ic_4 \theta^{\mu\nu}\theta^{\ka\la}\pat_\mu A^0_\ka
\pat_\nu A^0_\la \psi^0 ,\nn\\
\psi^{2, c_5}&=& c_5 \theta^{\mu\nu}\theta^{\ka\la}\big(i\{\pat_\mu A^0_\nu, A^0_\ka  A^0_\la \}\psi^0 +2 A^0_
\mu A^0_\nu A^0_\ka A^0_\la \psi^0 \big) ,\\
\psi^{2, c_6}&=& c_6 \theta^{\mu\nu}\theta^{\ka\la}2 A^0_
\mu A^0_\nu A^0_\ka A^0_\la \psi^0\nn,
\end{eqnarray} 
and for the gauge potential
\begin{eqnarray}
\label{fr28}
A^{2, c_4}_\rho &=& ic_4 \theta^{\mu\nu}\theta^{\ka\la}\big(\p_\rho(\pat_\mu A^0_\ka
\pat_\nu A^0_\la) -i\lb A^0_\rho,(\pat_\mu A^0_\ka
\pat_\nu A^0_\la)\rb\big),\nn\\
A^{2, c_5}_\rho&=& c_5 \theta^{\mu\nu}\theta^{\ka\la}\big(\p_\rho\{\pat_\mu
A^0_\nu, A^0_\ka  A^0_\la \}-i\lb A^0_\rho,\{\pat_\mu
A^0_\nu, A^0_\ka  A^0_\la \}\rb\big) ,\\
A^{2, c_6}_\rho&=& c_6 \theta^{\mu\nu}\theta^{\ka\la}2 \big(\p_\rho(A^0_
\mu A^0_\nu A^0_\ka A^0_\la)-i\lb A^0_\rho,A^0_
\mu A^0_\nu A^0_\ka A^0_\la\rb\big) .
\end{eqnarray} 
One can show that all actions built from
these fields are
identically zero. Although the list of ambiguities at second order
$\La_\al^{2, c_4}$, $\La_\al^{2, c_5}$ and $\La_\al^{2, c_6}$ is not
exhaustive, we are lead to  believe that such
ambiguities in $\La_\al$ are altogether irrelevant from the point of view of 
actions. We may concentrate on field redefinitions. 

Next we investigate in which sense the first order ambiguities (field redefinitions)
proportional to $c_2$ resp. $c_3$ reappear as
additional ambiguities  at second order. For example the ambiguity
parametrised by $c_2$: $\psi^{1,c_2}=c_2
\theta^{\mu\nu}F^0_{\mu\nu}\psi^0$ leads to the following $c_2$ parametrised Lagrangians and actions in second order:
\begin{eqnarray}
\label{fr21}
\mathcal{L}^{2,c_2}_{\textrm{\tiny MF}}&=&\frac{ic_2}{2}\theta^{\mu\nu}\theta^{\ka\la}\big(\mathcal{D}^0_\ka(\overline{\psi^0}F^0_{\mu\nu})\mathcal{D}^0_\la\psi^0 +\mathcal{D}^0_\ka\overline{\psi^0}\mathcal{D}^0_\la( F^0_{\mu\nu}\psi^0)\big)+c_2^2\theta^{\mu\nu}\theta^{\ka\la}\overline{\psi^0}F^0_{\mu\nu}F^0_{\ka\la}\psi^0 ,\nn\\
\mathcal{S}^{2,c_2}_{\textrm{\tiny MF}}&=&(c_2^2-\frac{c_2}{2})\theta^{\mu\nu}\theta^{\ka\la}\int \textrm{d}x \hspace{2mm}\overline{\psi^0}F^0_{\mu\nu}F^0_{\ka\la}\psi^0, \nn\\
\mathcal{L}^{2,c_2}_{\textrm{\tiny MCF}}&=&-\frac{c_2}{2}\theta^{\mu\nu}\theta^{\ka\la}\Big(\mathcal{D}^0_\ka(\overline{\psi^0}F^0_{\mu\nu})\ga^\rho\mathcal{D}^0_\la\mathcal{D}^0_\rho\psi^0 +\mathcal{D}^0_\ka\overline{\psi^0}\ga^\rho\mathcal{D}^0_\la(\mathcal{D}^0_\rho( F^0_{\mu\nu}\psi^0))\nn\\
&&\qquad\quad-i\overline{\psi^0}\ga^\rho F^0_{\mu\nu} F^0_{\ka\rho}\mathcal{D}^0_\la\psi^0 -i\overline{\psi^0}\ga^\rho F^0_{\ka\rho}\mathcal{D}^0_\la( F^0_{\mu\nu}\psi^0)\Big)\\
&&+ic_2^2\theta^{\mu\nu}\theta^{\ka\la}\overline{\psi^0}F^0_{\mu\nu}\ga^\rho\mathcal{D}^0_\rho (F^0_{\ka\la}\psi^0), \nn\\
\mathcal{S}^{2,c_2}_{\textrm{\tiny
    MCF}}&=&i\theta^{\mu\nu}\theta^{\ka\la}\int \textrm{d}x
\Big((c_2^2-\frac{c_2}{4})\overline{\psi^0}\ga^\rho
F^0_{\mu\nu}\mathcal{D}^0_\rho(F^0_{\ka\la}\psi^0)-\frac{c_2}{4}\overline{\psi^0}\ga^\rho
F^0_{\mu\nu} F^0_{\ka\la}\mathcal{D}^0_\rho\psi^0\nn\\
&&\qquad\qquad+\frac{c_2}{2}\overline{\psi^0}\ga^\rho F^0_{\mu\nu} F^0_{\ka\rho}\mathcal{D}^0_\la\psi^0 +\frac{c_2}{2}\overline{\psi^0}\ga^\rho F^0_{\ka\rho}\mathcal{D}^0_\la( F^0_{\mu\nu}\psi^0)\Big). \nn
\end{eqnarray} 
The gauge potential ambiguity proportional to $c_3$
leads in second order to
\begin{equation}
\label{fr22}
A^{2,c_3}_\rho=c_3 \frac{1}{2}\theta^{\mu\nu}\theta^{\ka\la}\Big(\{\pat_\mu(\mathcal{D}^0_\rho F^0_{\ka\la}), A^0_\nu\}-\frac{i}{2}\{\lb A^0_\mu, (\mathcal{D}^0_\rho F^0_{\ka\la})\rb, A^0_\nu\}\Big).
\end{equation} 
For the minimally coupled fermion action this means in second order:
\begin{eqnarray}
\label{fr23}
\mathcal{L}^{2,c_3}_{\textrm{\tiny MCF}}&=&\frac{ic_3}{2}\theta^{\mu\nu}\theta^{\ka\la}\big(\mathcal{D}^0_\ka\overline{\psi^0}\ga^\rho\mathcal{D}^0_\la ((\mathcal{D}^0_\rho F^0_{\mu\nu})\psi^0) +\overline{\psi^0}\ga^\rho(\mathcal{D}^0_\ka\mathcal{D}^0_\rho F^0_{\mu\nu})\mathcal{D}^0_\la\psi^0\big),\\
\mathcal{S}^{2,c_3}_{\textrm{\tiny MCF}}&=&c_3\theta^{\mu\nu}\theta^{\ka\la}\int \textrm{d}x \big(-\frac{1}{4}\overline{\psi^0}\ga^\rho F^0_{\mu\nu}(\mathcal{D}^0_\rho F^0_{\ka\la})\psi^0+\frac{i}{2}\overline{\psi^0}\ga^\rho(\mathcal{D}^0_\ka\mathcal{D}^0_\rho F^0_{\mu\nu})\mathcal{D}^0_\la\psi^0\big). \nn
\end{eqnarray} 
In addition we get a mixed term in $c_2$ and $c_3$
\begin{equation}
\label{fr23rrr}
\mathcal{S}^{2,c_2,c_3}_{\textrm{\tiny MCF}}=c_2
c_3\theta^{\mu\nu}\theta^{\ka\la}\int \textrm{d}x
\overline{\psi^0}\ga^\rho \{F^0_{\mu\nu}, (\mathcal{D}^0_\rho F^0_{\ka\la})\}\psi^0. 
\end{equation} 
Obviously, the field redefinitions at first order give additional
action terms at second order. We try to understand which terms are
physically relevant and compare with the 
fermionic actions at second order, (\ref{ac13a}) and (\ref{ac14a}):
\begin{eqnarray}
\label{fr29}
&&\int\textrm{d}x\hspace{1mm} \overline{\psi}\star(i\gamma^\rho
 {\cal{D}}_\rho-m)\psi|_{{\cal O}(\hbar^2)}=\nn\\
&&\quad \theta^{\mu\nu}\theta^{\ka\la}\int\textrm{d}x\hspace{1mm}\Big(
\frac{1}{32} \overline{\psi^0}  F^0_{\ka\la} F^0_{\mu\nu}(i\gamma^\rho
 {\cal{D}}^0_\rho-m)\psi^0-\frac{1}{8}\overline{\psi^0}
 F^0_{\ka\mu}  F^0_{\la\nu}(i\gamma^\rho{\cal{D}}^0_\rho-m)\psi^0
\nn\\
&&\qquad\qquad+\frac{i}{8}\overline{\psi^0}({\cal{D}}^0_\mu F^0_{\nu\ka})
 {\cal{D}}^0_\la(i\gamma^\rho{\cal{D}}^0_\rho-m)\psi^0-\frac{i}{8}\overline{\psi^0}\gamma^\rho
F^0_{\mu\ka}F^0_{\la\rho}{\cal{D}}^0_\nu\psi^0\\
&&\qquad\qquad-\frac{i}{4}\overline{\psi^0}\gamma^\rho
F^0_{\mu\rho}F^0_{\nu\ka}{\cal{D}}^0_\la\psi^0-\frac{1}{4}\overline{\psi^0}\gamma^\rho({\cal{D}}^0_\mu F^0_{\ka\rho}
) {\cal{D}}^0_\nu{\cal{D}}^0_\la\psi^0-\frac{i}{8}\overline{\psi^0}\gamma^\rho
 F^0_{\mu\nu}  F^0_{\ka\rho}{\cal{D}}^0_\la\psi^0\Big).\nn\end{eqnarray}

Since $c_2$ and $c_3$ are free parameters, we cannot use them to argue
 that some of the second-order action terms  in (\ref{fr29}) are physically
 irrelevant. But to this end we can investigate new second order
 field redefinitions:
\begin{eqnarray}
\label{fr30}
\psi^{2,c_{7}}&=&c_{7}\theta^{\mu\nu}\theta^{\ka\la} F^0_{\mu\nu}
F^0_{\ka\la}\psi^0,\nn\\
\psi^{2,c_{8}}&=&c_{8}\theta^{\mu\nu}\theta^{\ka\la} F^0_{\ka\mu}
F^0_{\la\nu}\psi^0,\\
\psi^{2,c_{9}}&=&ic_{9}\theta^{\mu\nu}\theta^{\ka\la}(\mathcal{D}^0_\mu F^0_{\nu\ka})\mathcal{D}^0_\la\psi^0.\nn
\end{eqnarray}
These lead to the following additional actions:
{\small \begin{eqnarray}
\label{fr31}
\mathcal{S}^{2,c_{7}}_{\textrm{\tiny MCF, MF}}&=&c_{7}\theta^{\mu\nu}\theta^{\ka\la}\int\textrm{d}x\hspace{1mm}\big(2 \overline{\psi^0}F^0_{\mu\nu}
F^0_{\ka\la}(i\gamma^\rho
 \mathcal{D}^0_\rho-m)\psi^0+i
 \overline{\psi^0}\gamma^\rho\{\mathcal{D}^0_\rho F^0_{\mu\nu},F^0_{\ka\la}\}\psi^0\big),\nn\\
\mathcal{S}^{2,c_{8}}_{\textrm{\tiny MCF, MF}}&=&c_{8}\theta^{\mu\nu}\theta^{\ka\la}\int\textrm{d}x\hspace{1mm}\big( 2\overline{\psi^0}F^0_{\ka\mu}
F^0_{\la\nu}(i\gamma^\rho
 \mathcal{D}^0_\rho-m)\psi^0+i
 \overline{\psi^0}\gamma^\rho\{\mathcal{D}^0_\rho F^0_{\ka\mu},F^0_{\la\nu}\}\psi^0\big),\\
\mathcal{S}^{2,c_{9}}_{\textrm{MCF,
    MF}}&=&c_{9}\theta^{\mu\nu}\theta^{\ka\la}\int\textrm{d}x\hspace{1mm}\big( 2i\overline{\psi^0}(\mathcal{D}^0_\ka F^0_{\la\mu})\mathcal{D}^0_\nu(i\ga^\rho \mathcal{D}^0_\rho-m)\psi^0-\overline{\psi^0}\ga^\rho(\mathcal{D}^0_\rho\mathcal{D}^0_\ka F^0_{\la\mu})\mathcal{D}^0_\nu\psi^0\big).\nn
\end{eqnarray}}
The ambiguities (\ref{fr30}) show that the first
three terms in  (\ref{fr29}), especially all fermionic mass terms, are
of the type of a field redefinition and therefore should vanish. The
additional terms in the action introduced in (\ref{fr31}) can be
cancelled by additional redefinitions of the gauge potential,
e.g.
$A^{2,c_{10}}_\rho=c_{10}\theta^{\mu\nu}\theta^{\ka\la}\{\mathcal{D}^0_\rho
F^0_{\mu\nu},F^0_{\ka\la}\}$ etc.  In
contrast, the last four terms in  (\ref{fr29}) are not affected, they
are not of the type of  field redefinitions. Therefore the only
physically relevant terms in the fermionic action to second order are
\begin{eqnarray}
\label{fr30a}
&&\int\textrm{d}x\hspace{1mm} \overline{\psi}\star(i\gamma^\rho
 {\cal{D}}_\rho-m)\psi|_{{\cal O}(\hbar^2)}=\theta^{\mu\nu}\theta^{\ka\la}\int\textrm{d}x\hspace{1mm}\Big(-\frac{i}{8}\overline{\psi^0}\gamma^\rho
F^0_{\mu\ka}F^0_{\la\rho}{\cal{D}}^0_\nu\psi^0\\
&&\qquad\qquad -\frac{i}{4}\overline{\psi^0}\gamma^\rho
F^0_{\mu\rho}F^0_{\nu\ka}{\cal{D}}^0_\la\psi^0-\frac{1}{4}\overline{\psi^0}\gamma^\rho({\cal{D}}^0_\mu F^0_{\ka\rho}
) {\cal{D}}^0_\nu{\cal{D}}^0_\la\psi^0-\frac{i}{8}\overline{\psi^0}\gamma^\rho
 F^0_{\mu\nu}  F^0_{\ka\rho}{\cal{D}}^0_\la\psi^0\Big).\nn
\end{eqnarray}
The Yang-Mills action is not affected by these field redefitions\footnote{In a recent article \cite{banerjee1} it has been 
 argued that the ambiguity freedom of the Maxwell action can be shown 
to all orders; since it has been shown \cite{banerjee2} that both Abelian and non-Abelian source terms remain ambiguity-free, we assume that our result for NC Yang-Mills will be valid to all orders.} ,
since the field stength corresponding e.g. to $A^{2,c_{10}}_\rho$,
again only results in a commutator term in the action. 

%
%
%
%
%

\section{Conclusion}
\label{F}

From the point of view of the theory of consistent deformations
(cp. \cite{brandt4}), the enveloping algebra is a nontrivial
deformation of type 1. This means that although the
gauge transformation is trivially deformed
($\de_{\La_\al}\equiv\de_\al$), the deformation is non-trivial, it cannot
only be obtained via field redefinitions. Indeed we have seen that both the Yang-Mills action and the
fermionic interaction term  (\ref{fr19a}) are nontrivial at first and
second order.

\section*{Acknowledgements}
I am grateful to Marija Dimitrijevi\'c, Branislav Jur\v co and Julius Wess for many 
fruitful discussions, contributing to the results of this note. Similarly I am grateful to Josip Trampeti\'c for making valuable 
suggestions for improvement of the manuscript and to Marija for careful proof-reading.


\end{document}